\batchmode
\makeatletter
\def\input@path{{\string"C:/Users/rushik/Desktop/New folder/\string"}}
\makeatother
\documentclass[english,9pt,technote]{IEEEtran}
\usepackage[T1]{fontenc}
\usepackage[latin9]{inputenc}
\usepackage{verbatim}
\usepackage{units}
\usepackage{amsmath}
\usepackage{amsthm}
\usepackage{amssymb}
\usepackage{graphicx}
\usepackage{esint}

\makeatletter
\theoremstyle{definition}
\newtheorem{assumption}{Assumption}
  \theoremstyle{remark}
  \newtheorem{rem}{\protect\remarkname}
  \theoremstyle{plain}
  \newtheorem{thm}{\protect\theoremname}

\usepackage{cite}
\IEEEoverridecommandlockouts

\makeatother

\usepackage{babel}
\providecommand{\remarkname}{Remark}
\providecommand{\theoremname}{Theorem}

\begin{document}

\title{Online Approximate Optimal Station Keeping of a Marine Craft in the
Presence of a Current\thanks{Patrick Walters, Rushikesh Kamalapurkar, and Warren E. Dixon are with
the Department of Mechanical and Aerospace Engineering, University
of Florida, Gainesville, FL, USA (Email: \{walters8, rkamalapurkar,
wdixon\}@ufl.edu). Forrest Voight and Eric M. Schwartz are with the
Department of Electrical and Computer Engineering, University of Florida,
Gainesville, FL, USA (Email: \{forrestv, ems\}@ufl.edu).}\thanks{This research is supported in part by NSF award numbers 0901491, 1161260,
1217908, ONR grant number N00014-13-1-0151, and a contract with the
AFRL Mathematical Modeling and Optimization Institute. Any opinions,
findings and conclusions or recommendations expressed in this material
are those of the authors and do not necessarily reflect the views
of the sponsoring agency.}}

\author{Patrick Walters, Rushikesh Kamalapurkar, Forrest Voight, Eric M.
Schwartz, and Warren E. Dixon}
\maketitle
\begin{abstract}
Online approximation of the optimal station keeping strategy for a
fully actuated six degrees-of-freedom marine craft subject to an irrotational
ocean current is considered. An approximate solution to the optimal
control problem is obtained using an adaptive dynamic programming
technique. The hydrodynamic drift dynamics of the dynamic model are
assumed to be unknown; therefore, a concurrent learning-based system
identifier is developed to identify the unknown model parameters.
The identified model is used to implement an adaptive model-based
reinforcement learning technique to estimate the unknown value function.
The developed policy guarantees uniformly ultimately bounded convergence
of the vehicle to the desired station and uniformly ultimately bounded
convergence of the approximated policies to the optimal polices without
the requirement of persistence of excitation. The developed strategy
is validated using an autonomous underwater vehicle, where the three
degrees-of-freedom in the horizontal plane are regulated. The experiments
are conducted in a second-magnitude spring located in central Florida.
\end{abstract}


\section{Introduction}

Marine craft, which include ships, floating platforms, autonomous
underwater vehicles (AUV), etc, play a vital role in commercial, military
and recreational objectives. Marine craft are often required to remain
on a station for an extended period of time, e.g., floating oil platforms,
support vessels, and AUVs acting as a communication link for multiple
vehicles or persistent environmental monitors. The success of the
vehicle often relies on the vehicle's ability to hold a precise station
(e.g., station keeping near structures or underwater features). The
cost of holding that station is correlated to the energy expended
for propulsion through consumption of fuel and wear on mechanical
systems, especially when station keeping in environments with a persistent
current. Therefore, by reducing the energy expended for station keeping
objectives, the cost of holding a station can be reduced.

Precise station keeping of a marine craft is challenging because of
nonlinearities in the dynamics of the vehicle. A survey of station
keeping for surface vessels can be found in \cite{Soerensen2011}.
Common approaches employed to control a marine craft include robust
and adaptive control methods \cite{Fossen.Grovlen1998,Sebastian2007,Tannuri.Agostinho.ea2010,Fischer.Hughes.ea2014}.
These methods provide robustness to disturbances and/or model uncertainty;
however, they do not explicitly account for the cost of the control
effort. Motivated by the desire to balance energy expenditure and
the accuracy of the vehicle\textquoteright s station, approximate
optimal control methods are examined in this paper to minimize a user
defined cost function of the total control effort (energy expended)
and state error (station accuracy). Because of the difficulties associated
with finding closedform analytical solutions to optimal control problems
for marine craft, efforts such as \cite{Beard.Mclain1998} numerically
approximate the solution to the Hamilton-Jacobi-Bellman (HJB) equation
using an iterative application of Galerkin's method.

Various methods have been proposed to find an approximate solution
to the HJB equation. Adaptive dynamic programming (ADP) is one such
method where a solution to the HJB equation is approximated using
parametric function approximation techniques. ADP-based techniques
have been used to approximate optimal control policies for regulation
(e.g., \cite{Bhasin.Kamalapurkar.ea2013a,Vrabie2009,Vamvoudakis2010,Modares.Lewis.ea2013a,Modares.Lewis.ea2014})
of general nonlinear systems. Efforts in \cite{Johnson2011a} and
\cite{Vamvoudakis2010a} present ADP-based solutions to the Hamilton-Jacobi-Isaacs
equation that yield an approximate optimal policy accounting for state-dependent
disturbances. However, these methods do not consider explicit time-varying
disturbances such as the dynamics that are introduced due to the presence
of current.

In this result, an optimal station keeping policy that captures the
desire to balance the need to accurately hold a station and the cost
of holding that station through a quadratic performance criterion
is generated for a fully actuated marine craft. The developed controller
differs from results such as \cite{Bhasin.Kamalapurkar.ea2013a} and
\cite{Vrabie2009} in that it tackles the challenges associated with
the introduction of a time-varying irrotational current. Since the
hydrodynamic parameters of a marine craft are often difficult to determine,
a concurrent learning system identifier is developed. As outlined
in \cite{Chowdhary.Yucelen.ea2012}, concurrent learning uses additional
information from recorded data to remove the persistence of excitation
requirement associated with traditional system identifiers. The proposed
model-based ADP method generates the optimal station keeping policy
using a combination of on-policy and off-policy data, eliminating
the need for physical exploration of the state space. A Lyapunov-based
stability analysis is presented which guarantees uniformly ultimately
bounded (UUB) convergence of the marine craft to its station and UUB
convergence of the approximated policy to the optimal policy. 

To illustrate the performance of the developed controller, an AUV
is used to collect experimental data. Specifically, the developed
strategy is implemented for planar regulation of an AUV near the vent
of a second-magnitude spring located in central Florida. The experimental
results demonstrate the developed method\textquoteright s ability
to simultaneously identify the unknown hydrodynamic parameters and
generate an approximate optimal policy using the identified model
in the presence of a current.

\section{Vehicle Model}

Consider the nonlinear equations of motion for a marine craft including
the effects of irrotational ocean current given in Section 7.5 of
\cite{Fossen2011} as
\begin{equation}
\dot{\eta}=J_{E}\left(\eta\right)\nu,\label{eq:kinematics}
\end{equation}
\begin{eqnarray}
M_{RB}\dot{\nu}+C_{RB}\left(\nu\right)\nu+M_{A}\dot{\nu}_{r}+C_{A}\left(\nu_{r}\right)\nu_{r}\nonumber \\
+D_{A}\left(\nu_{r}\right)\nu_{r}+G\left(\eta\right) & = & \tau_{b}\label{eq:dynamics}
\end{eqnarray}
where $\nu\in\mathbb{R}^{n}$ is the body-fixed translational and
angular velocity vector, $\nu_{c}\in\mathbb{R}^{n}$ is the body-fixed
irrotational current velocity vector, $\nu_{r}=\nu-\nu_{c}$ is the
relative body-fixed translational and angular fluid velocity vector,
$\eta\in\mathbb{R}^{n}$ is the earth-fixed position and orientation
vector, $J_{E}:\mathbb{R}^{n}\rightarrow\mathbb{R}^{n\times n}$ is
the coordinate transformation between the body-fixed and earth-fixed
coordinates\footnote{The orientation of the vehicle may be represented as Euler angles,
quaternions, or angular rates. In this development, the use of Euler
angles is assumed, see Section 7.5 in \cite{Fossen2011} for details
regarding other representations.}, $M_{RB}\in\mathbb{R}^{n\times n}$ is the constant rigid body inertia
matrix, $C_{RB}:\mathbb{R}^{n}\rightarrow\mathbb{R}^{n\times n}$
is the rigid body centripetal and Coriolis matrix, $M_{A}\in\mathbb{R}^{n\times n}$
is the constant hydrodynamic added mass matrix, $C_{A}:\mathbb{R}^{n}\rightarrow\mathbb{R}^{n\times n}$
is the unknown hydrodynamic centripetal and Coriolis matrix, $D_{A}:\mathbb{R}^{n}\rightarrow\mathbb{R}^{n\times n}$
is the unknown hydrodynamic damping and friction matrix, $G:\mathbb{R}^{n}\rightarrow\mathbb{R}^{n}$
is the gravitational and buoyancy force and moment vector, and $\tau_{b}\in\mathbb{R}^{n}$
is the body-fixed force and moment control input.

In the case of a three degree-of-freedom (DOF) planar model with orientation
represented as Euler angles, the state vectors in $\eqref{eq:kinematics}$
and (\ref{eq:dynamics}) are further defined as
\[
\eta\triangleq\left[\begin{array}{ccc}
x & y & \psi\end{array}\right]^{T},
\]
\[
\nu\triangleq\left[\begin{array}{ccc}
u & v & r\end{array}\right]^{T},
\]
where $x$, $y\in\mathbb{R}$, are the earth-fixed position vector
components of the center of mass, $\psi\in\mathbb{R}$ represents
the yaw angle, $u$, $v\in\mathbb{R}$ are the body-fixed translational
velocities, and $r\in\mathbb{R}$ is the body-fixed angular velocity.
The irrotational current vector is defined as 
\[
\nu_{c}\triangleq\left[\begin{array}{ccc}
u_{c} & v_{c} & 0\end{array}\right]^{T},
\]
where $u_{c}$, $v_{c}\in\mathbb{R}$ are the body-fixed current translational
velocities. The coordinate transformation $J_{E}\left(\eta\right)$
is given as
\[
J_{E}\left(\eta\right)=\left[\begin{array}{ccc}
\cos\left(\psi\right) & -\sin\left(\psi\right) & 0\\
\sin\left(\psi\right) & \cos\left(\psi\right) & 0\\
0 & 0 & 1
\end{array}\right].
\]
\begin{assumption}
\label{thm:neutral}The marine craft is neutrally buoyant if submerged
and the center of gravity is located vertically below the center of
buoyancy on the $z$ axis if the vehicle model includes roll and pitch\footnote{This assumption simplifies the subsequent analysis and can often be
met by trimming the vehicle. For marine craft where this assumption
cannot be met, an additional term may be added to the controller,
similar to how terms dependent on the irrotational current are handled.}. 
\end{assumption}

\section{System Identifier}

Since the hydrodynamic effects pertaining to a specific marine craft
may be unknown, an online system identifier is developed for the vehicle
drift dynamics. Consider the control affine form of the vehicle model,
\begin{equation}
\dot{\zeta}=Y\left(\zeta,\nu_{c}\right)\theta+f_{0}\left(\zeta,\dot{\nu}_{c}\right)+g\tau_{b},\label{eq:dyn}
\end{equation}
where $\zeta\triangleq\left[\begin{array}{cc}
\eta & \nu\end{array}\right]^{T}\in\mathbb{R}^{2n}$ is the state vector. The unknown hydrodynamics are linear-in-the-parameters
with $p$ unknown parameters where $Y:\mathbb{R}^{2n}\times\mathbb{R}^{n}\rightarrow\mathbb{R}^{2n\times p}$
is the regression matrix and $\theta\in\mathbb{R}^{p}$ is the vector
of unknown parameters. The unknown hydrodynamic effects are modeled
as 
\[
Y\left(\zeta,\nu_{c}\right)\theta=\left[\begin{array}{c}
0\\
-M^{-1}C_{A}\left(\nu_{r}\right)\nu_{r}-M^{-1}D_{A}\left(\nu_{r}\right)\nu_{r}
\end{array}\right],
\]
and known rigid body drift dynamics $f_{0}:\mathbb{R}^{2n}\times\mathbb{R}^{n}\rightarrow\mathbb{R}^{2n}$
are modeled as 
\begin{multline*}
f_{0}\left(\zeta,\dot{\nu}_{c}\right)=\\
\left[\begin{array}{c}
J_{E}\left(\eta\right)\nu\\
M^{-1}M_{A}\dot{\nu}_{c}-M^{-1}C_{RB}\left(\nu\right)\nu-M^{-1}G\left(\eta\right)
\end{array}\right],
\end{multline*}
where $M\triangleq M_{RB}+M_{A}$, and the body-fixed current velocity
$\nu_{c}$, and acceleration $\dot{\nu}_{c}$ are assumed to be measurable\footnote{The body-fixed current velocity $\nu_{c}$ may be trivially measured
using sensors commonly found on marine craft, such as a Doppler velocity
log, while the current acceleration $\dot{\nu}_{c}$ may be determined
using numerical differentiation and smoothing.}. The known constant control effectiveness matrix $g\in\mathbb{R}^{2n\times n}$
is defined as 
\[
g\triangleq\left[\begin{array}{c}
0\\
M^{-1}
\end{array}\right].
\]
An identifier is designed as
\begin{equation}
\dot{\hat{\zeta}}=Y\left(\zeta,\nu_{c}\right)\hat{\theta}+f_{0}\left(\zeta,\dot{\nu}_{c}\right)+g\tau_{b}+k_{\zeta}\tilde{\zeta},\label{eq:ident}
\end{equation}
where $\tilde{\zeta}\triangleq\zeta-\hat{\zeta}$ is the measurable
state estimation error, and $k_{\zeta}\in\mathbb{R}^{2n\times2n}$
is a constant positive definite, diagonal gain matrix. Subtracting
(\ref{eq:ident}) from (\ref{eq:dyn}), yields
\[
\dot{\tilde{\zeta}}=Y\left(\zeta,\nu_{c}\right)\tilde{\theta}-k_{\zeta}\tilde{\zeta},
\]
where $\tilde{\theta}\triangleq\theta-\hat{\theta}$ is the parameter
identification error.

\subsection{Parameter Update}

Traditional adaptive control techniques require persistence of excitation
to ensure the parameter estimates $\hat{\theta}$ converge to their
true values $\theta$ (cf. \cite{Sastry1989} and \cite{Ioannou1996}).
Persistence of excitation often requires an excitation signal to be
applied to the vehicle's input resulting in unwanted deviations in
the vehicle state. These deviations are often in opposition to the
vehicle's control objectives. Alternatively, a concurrent learning-based
system identifier can be developed (cf. \cite{Chowdhary.Yucelen.ea2012}
and \cite{Chowdhary.Johnson2011a}). The concurrent learning-based
system identifier relaxes the persistence of excitation requirement
through the use of a prerecorded history stack of state-action pairs\footnote{In this development, it is assumed that a data set of state-action
pairs is available a priori. Experiments to collect state-action pairs
do not necessarily need to be conducted in the presence of a current
(e.g. the data may be collected in a pool). Since the current affects
the dynamics only through the $\nu_{r}$ terms, data that is sufficiently
rich and satisfies Assumption \ref{thm:concurrent-1} may be collected
by merely exploring the $\zeta$ state space. Note, this is the reason
the body-fixed current $\nu_{c}$ and acceleration $\dot{\nu}_{c}$
are not considered a part of the state. If state-action data is not
available for the given system then it is possible to build the history
stack in real-time and the details of that development can be found
in Appendix A of \cite{Kamalapurkar2014}.}.
\begin{assumption}
\label{thm:concurrent-1}There exists a prerecorded data set of sampled
data points $\left\{ \zeta_{j},\nu_{cj},\dot{\nu}_{cj},\tau_{bj}\in\chi|j=1,2,\ldots,M\right\} $
with a numerically calculated state derivatives $\dot{\bar{\zeta}}_{j}$
at each recorded state-action pair such that $\forall t\in\left[0,\infty\right)$,
\begin{equation}
\mathrm{rank}\left(\sum_{j=1}^{M}Y_{j}^{T}Y_{j}\right)=p,\label{eq:rank_cond-1}
\end{equation}
\[
\left\Vert \dot{\bar{\zeta}}_{j}-\dot{\zeta}_{j}\right\Vert <\bar{d},\forall j,
\]
where $Y_{j}\triangleq Y\left(\zeta_{j},\nu_{cj}\right)$, $f_{0j}\triangleq f_{0}\left(\zeta_{j}\right)$,
$\dot{\zeta}_{j}=Y_{j}\theta+f_{0j}+g\tau_{bj}$, and $\bar{d}\in\left[0,\infty\right)$
is a constant. 
\end{assumption}

The parameter estimate update law is given as
\begin{equation}
\dot{\hat{\theta}}=\Gamma_{\theta}Y\left(\zeta,\nu_{c}\right)^{T}\tilde{\zeta}+\Gamma_{\theta}k_{\theta}\sum_{j=1}^{M}Y_{j}^{T}\left(\dot{\bar{\zeta}}_{j}-f_{0j}-g\tau_{bj}-Y_{j}\hat{\theta}\right),\label{eq:para_update}
\end{equation}
where $\Gamma_{\theta}$ is a positive definite, diagonal gain matrix,
and $k_{\theta}$ is a positive, scalar gain matrix. To facilitate
the stability analysis, the parameter estimate update law is expressed
in the advantageous form%
\[
\dot{\hat{\theta}}=\Gamma_{\theta}Y\left(\zeta,\nu_{c}\right)^{T}\tilde{\zeta}+\Gamma_{\theta}k_{\theta}\sum_{j=1}^{M}Y_{j}^{T}\left(Y_{j}\tilde{\theta}+d_{j}\right),
\]
where $d_{j}=\dot{\bar{\zeta}}_{j}-\dot{\zeta}_{j}$.
\begin{rem}
The update law in (\ref{eq:para_update}) does not require \emph{instantaneous}\textit{\emph{
measurement of acceleration. Acceleration only needs to be computed
at the }}\textit{past}\textit{\emph{ time instances when the data
points $\left(\zeta_{j},\nu_{cj},\dot{\nu}_{cj},\tau_{bj}\right)$
were recorded. Acceleration at a past time instance $t^{*}$ can be
accurately computed by recording position and velocity signals over
a time interval that contains $t^{*}$ in its interior and using noncausal
estimation methods such as optimal fixed-point smoothing \cite[p. 170]{Gelb1974}.}}
\end{rem}

\subsection{Convergence Analysis}

Consider the candidate Lyapunov function $V_{P}:\mathbb{R}^{2n+p}\times\left[0,\infty\right)$
given as
\begin{equation}
V_{P}\left(Z_{P}\right)=\frac{1}{2}\tilde{\zeta}^{T}\tilde{\zeta}+\frac{1}{2}\tilde{\theta}^{T}\Gamma_{\theta}^{-1}\tilde{\theta},\label{eq:id_lyp}
\end{equation}
where $Z_{P}\triangleq\left[\begin{array}{cc}
\tilde{\zeta}^{T} & \tilde{\theta}^{T}\end{array}\right]$. The candidate Lyapunov function can be bounded as
\begin{equation}
\frac{1}{2}\min\left\{ 1,\underline{\gamma_{\theta}}\right\} \left\Vert Z_{P}\right\Vert ^{2}\leq V_{P}\left(Z_{P}\right)\leq\frac{1}{2}\max\left\{ 1,\overline{\gamma_{\theta}}\right\} \left\Vert Z_{P}\right\Vert ^{2}\label{eq:id_classK}
\end{equation}
where $\underline{\gamma_{\theta}},\overline{\gamma_{\theta}}$ are
the minimum and maximum eigenvalues of $\Gamma_{\theta}$, respectively.

The time derivative of the candidate Lyapunov function in (\ref{eq:id_lyp})
is%
{} 
\[
\dot{V}_{P}=-\tilde{\zeta}^{T}k_{\zeta}\tilde{\zeta}-k_{\theta}\tilde{\theta}^{T}\sum_{j=1}^{M}Y_{j}^{T}Y_{j}\tilde{\theta}-k_{\theta}\tilde{\theta}^{T}\sum_{j=1}^{M}Y_{j}^{T}d_{j}.
\]
The time derivative may be upper bounded by
\begin{equation}
\dot{V}_{P}\leq-\underline{k_{\zeta}}\left\Vert \tilde{\zeta}\right\Vert ^{2}-k_{\theta}\underline{y}\left\Vert \tilde{\theta}\right\Vert ^{2}+k_{\theta}d_{\theta}\left\Vert \tilde{\theta}\right\Vert ,\label{eq:id_upperBound}
\end{equation}
where $\underline{k_{\zeta}},\underline{y}$ are the minimum eigenvalues
of $k_{\zeta}$ and $\sum_{j=1}^{M}Y_{j}^{T}Y_{j}$, respectively,
and $d_{\theta}=\bar{d}\sum_{j=1}^{M}\left\Vert Y_{j}\right\Vert $.
Completing the squares, (\ref{eq:id_upperBound}) may be upper bounded
by%
\[
\dot{V}_{P}\leq-\underline{k_{\zeta}}\left\Vert \tilde{\zeta}\right\Vert ^{2}-\frac{k_{\theta}\underline{y}}{2}\left\Vert \tilde{\theta}\right\Vert ^{2}+\frac{k_{\theta}d_{\theta}^{2}}{2\underline{y}},
\]
which may be further upper bounded by
\begin{equation}
\dot{V}_{P}\leq-\alpha_{P}\left\Vert Z_{P}\right\Vert ^{2},\forall\left\Vert Z_{P}\right\Vert \geq K_{P}>0,\label{eq:id_uub}
\end{equation}
where $\alpha_{P}\triangleq\frac{1}{2}\min\left\{ 2\underline{k_{\zeta}},k_{\theta}\underline{y}\right\} $
and $K_{P}\triangleq\sqrt{\frac{k_{\theta}d_{\theta}^{2}}{2\alpha_{P}\underline{y}}}$.
Using (\ref{eq:id_classK}) and (\ref{eq:id_uub}), $\tilde{\zeta}$
and $\tilde{\theta}$ can be shown to exponentially decay to a ultimate
bound as $t\rightarrow\infty$. The ultimate bound may be made arbitrarily
small depending on the selection of the gains $k_{\zeta}$ and $k_{\theta}$.

\section{Problem Formulation}

\subsection{Residual Model}

The presence of a time-varying irrotational current yields unique
challenges in the formulation of the optimal regulation problem. Since
the current renders the system non-autonomous, a residual model that
does not include the effects of the irrotational current is introduced.
The residual model is used in the development of the optimal control
problem in place of the original model. A disadvantage of this approach
is that the optimal policy is developed for the current-free model\footnote{To the author's knowledge, there is no method to generate a policy
with time-varying inputs (e.g., time-varying irrotational current)
that guarantees optimally and stability. }. In the case where the earth-fixed current is constant, the effects
of the current may be included in the development of the optimal control
problem as detailed in Appendix \ref{sec:Appendix_conCurr}.

The residual model can be written in a control affine form as
\begin{equation}
\dot{\zeta}=Y_{res}\left(\zeta\right)\theta+f_{0_{res}}\left(\zeta\right)+gu,\label{eq:adp_dyn}
\end{equation}
where the unknown hydrodynamics are linear-in-the-parameters with
$p$ unknown parameters where $Y_{res}:\mathbb{R}^{2n}\rightarrow\mathbb{R}^{2n\times p}$
is a regression matrix, the function $f_{0_{res}}:\mathbb{R}^{2n}\rightarrow\mathbb{R}^{2n}$
is the known portion of the dynamics, and $u\in\mathbb{R}^{n}$ is
the control vector. The drift dynamics, defined as $f_{res}\left(\zeta\right)=Y_{res}\left(\zeta\right)\theta+f_{0_{res}}\left(\zeta\right)$,
can be shown to satisfy $f_{res}\left(0\right)=0$ when Assumption
\ref{thm:neutral} is satisfied.

The drift dynamics in (\ref{eq:adp_dyn}) are modeled as%
\[
Y_{res}\left(\zeta\right)\theta=\left[\begin{array}{c}
0\\
-M^{-1}C_{A}\left(\nu\right)\nu-M^{-1}D\left(\nu\right)\nu
\end{array}\right],
\]
\begin{equation}
f_{0_{res}}\left(\zeta\right)=\left[\begin{array}{c}
J_{E}\nu\\
-M^{-1}C_{RB}\left(\nu\right)\nu-M^{-1}G\left(\eta\right)
\end{array}\right],\label{eq:adp_dynamics_varCur}
\end{equation}
and the virtual control vector $u$ is defined as
\begin{equation}
u=\tau_{b}-\tau_{c}\left(\zeta,\nu_{c},\dot{\nu}_{c}\right),\label{eq:virtual_control}
\end{equation}
where $\tau_{c}:\mathbb{R}^{2n}\times\mathbb{R}^{n}\times\mathbb{R}^{n}\rightarrow\mathbb{R}^{n}$
is a feedforward term to compensate for the effect of the variable
current, which includes cross-terms generated by the introduction
of the residual dynamics and is given as 
\begin{multline*}
\tau_{c}\left(\zeta,\nu_{c},\dot{\nu}_{c}\right)=C_{A}\left(\nu_{r}\right)\nu_{r}+D\left(\nu_{r}\right)\nu_{r}-M_{A}\dot{\nu}_{c}\\
-C_{A}\left(\nu\right)\nu-D\left(\nu\right)\nu.
\end{multline*}
The current feedforward term is represented in the advantageous form
\[
\tau_{c}\left(\zeta,\nu_{c},\dot{\nu}_{c}\right)=-M_{A}\dot{\nu}_{c}+Y_{c}\left(\zeta,\nu_{c}\right)\theta,
\]
where $Y_{c}:\mathbb{R}^{2n}\times\mathbb{R}^{n}\rightarrow\mathbb{R}^{2n\times p}$
is the regression matrix and
\[
Y_{c}\theta\left(\zeta,\nu_{c}\right)=C_{A}\left(\nu_{r}\right)\nu_{r}+D\left(\nu_{r}\right)\nu_{r}-C_{A}\left(\nu\right)\nu-D\left(\nu\right)\nu.
\]
Since the parameters are unknown, an approximation of the compensation
term $\tau_{c}$ given by
\begin{equation}
\hat{\tau}_{c}\left(\zeta,\nu_{c},\dot{\nu}_{c},\hat{\theta}\right)=-M_{A}\dot{\nu}_{c}+Y_{c}\hat{\theta}\label{eq:tau_c_approx}
\end{equation}
is implemented, and the approximation error is defined by 
\[
\tilde{\tau}_{c}\triangleq\tau_{c}-\hat{\tau}_{c}.
\]

\subsection{Nonlinear Optimal Regulation Problem }

The performance index for the optimal regulation problem is selected
as 
\begin{equation}
J\left(\zeta,u\right)=\intop_{0}^{\infty}r\left(\zeta\left(\tau\right),u\left(\tau\right)\right)d\tau,\label{eq:perform_index}
\end{equation}
where $r:\mathbb{R}^{2n}\rightarrow\left[0,\infty\right)$ is the
local cost defined as
\begin{equation}
r\left(\zeta,u\right)\triangleq\zeta^{T}Q\zeta+u^{T}Ru.\label{eq:local_cost}
\end{equation}
In $\eqref{eq:local_cost}$, $Q\in\mathbb{R}^{2n\times2n}$ , $R\in\mathbb{R}^{n\times n}$
are symmetric positive definite weighting matrices, and $u$ is the
virtual control vector. The matrix $Q$ has the property $\underline{q}\left\Vert \xi_{q}\right\Vert ^{2}\leq\xi_{q}^{T}Q\xi_{q}\leq\overline{q}\left\Vert \xi_{q}\right\Vert ^{2},\:\forall\xi_{q}\in\mathbb{R}^{2n}$
where $\underline{q}$ and $\overline{q}$ are positive constants.
The infinite-time scalar value function $V:\mathbb{R}^{2n}\rightarrow\left[0,\infty\right)$
for the optimal solution is written as
\begin{equation}
V\left(\zeta\right)=\underset{u}{\min}\intop_{0}^{\infty}r\left(\zeta\left(\tau\right),u\left(\tau\right)\right)d\tau.\label{eq:value}
\end{equation}

The objective of the optimal control problem is to find the optimal
policy $u^{*}:\mathbb{R}^{2n}\rightarrow\mathbb{R}^{n}$ that minimizes
the performance index $\eqref{eq:perform_index}$ subject to the dynamic
constraints in $\eqref{eq:adp_dyn}$. Assuming that a minimizing policy
exists and the value function is continuously differentiable, the
Hamiltonian $H:\mathbb{R}^{2n}\rightarrow\mathbb{R}$ is defined as
\begin{multline}
H\left(\zeta\right)\triangleq r\left(\zeta,u^{*}\left(\zeta\right)\right)\\
+\frac{\partial V\left(\zeta\right)}{\partial\zeta}\left(Y_{res}\left(\zeta\right)\theta+f_{0_{res}}\left(\zeta\right)+gu^{*}\left(\zeta\right)\right).\label{eq:ham}
\end{multline}
The HJB equation is given as \cite{Kirk2004}
\begin{equation}
0=\frac{\partial V\left(\zeta\right)}{\partial t}+H\left(\zeta\right),\label{eq:HJI}
\end{equation}
where $\frac{\partial V\left(\zeta\right)}{\partial t}=0$ since the
value function is not an explicit function of time. After substituting
$\eqref{eq:local_cost}$ into $\eqref{eq:HJI}$ , the optimal policy
is given by \cite{Kirk2004}
\begin{equation}
u^{*}\left(\zeta\right)=-\frac{1}{2}R^{-1}g^{T}\left(\frac{\partial V\left(\zeta\right)}{\partial\zeta}\right)^{T},\label{eq:u_opt}
\end{equation}

The analytical expression for the optimal controller in $\eqref{eq:u_opt}$
requires knowledge of the value function which is the solution to
the HJB equation in $\eqref{eq:HJI}$. The HJB equation is a partial
differential equation which is generally infeasible to solve; hence,
an approximate solution is sought.

\section{Approximate Policy}

The subsequent development is based on a neural network (NN) approximation
of the value function and optimal policy. Differing from previous
ADP literature with model uncertainty (e.g., \cite{Vrabie2009,Modares.Lewis.ea2013a,Modares.Lewis.ea2014})
that seeks a NN approximation using the integral form of the HJB,
the following development seeks a NN approximation using the differential
form. The differential form of the HJB coupled with the identified
model allows off-policy learning, which relaxes the persistence of
excitation condition previously required.

Over any compact domain $\chi\subset\mathbb{R}^{2n}$, the value function
$V:\mathbb{R}^{2n}\rightarrow\left[0,\infty\right)$ can be represented
by a single-layer NN with $l$ neurons as
\begin{equation}
V\left(\zeta\right)=W^{T}\sigma\left(\zeta\right)+\epsilon\left(\zeta\right),\label{eq:value_NN}
\end{equation}
where $W\in\mathbb{R}^{l}$ is the ideal weight vector bounded above
by a known positive constant, $\sigma:\mathbb{R}^{2n}\rightarrow\mathbb{R}^{l}$
is a bounded, continuously differentiable activation function, and
$\epsilon:\mathbb{R}^{2n}\rightarrow\mathbb{R}$ is the bounded, continuously
differential function reconstruction error. Using $\eqref{eq:u_opt}$
and $\eqref{eq:value_NN}$, the optimal policy can be represented
by
\begin{equation}
u^{*}\left(\zeta\right)=-\frac{1}{2}R^{-1}g^{T}\left(\sigma'\left(\zeta\right){}^{T}W+\epsilon'\left(\zeta\right){}^{T}\right),\label{eq:u_NN}
\end{equation}
where $\sigma':\mathbb{R}^{2n}\rightarrow\mathbb{R}^{l\times2n}$
and $\epsilon':\mathbb{R}^{2n}\rightarrow\mathbb{R}^{2n}$ are derivatives
with respect to the state. Based on $\eqref{eq:value_NN}$ and $\eqref{eq:u_NN}$,
NN approximations of the value function and the optimal policy are
defined as
\begin{equation}
\hat{V}\left(\zeta,\hat{W}_{c}\right)=\hat{W}_{c}^{T}\sigma\left(\zeta\right),\label{eq:V_approx}
\end{equation}
\begin{equation}
\hat{u}\left(\zeta,\hat{W}_{a}\right)=-\frac{1}{2}R^{-1}g^{T}\sigma'\left(\zeta\right)^{T}\hat{W}_{a},\label{eq:u_approx}
\end{equation}
where $\hat{W}_{c},\:\hat{W}_{a}\in\mathbb{R}^{l}$ are estimates
of the constant ideal weight vector $W$. The weight estimation errors
are defined as $\tilde{W}_{c}\triangleq W-\hat{W}_{c}$ and $\tilde{W}_{a}\triangleq W-\hat{W}_{a}$. 

Substituting (\ref{eq:adp_dyn}), $\eqref{eq:V_approx}$, and $\eqref{eq:u_approx}$
into $\eqref{eq:ham}$, the approximate Hamiltonian $\hat{H}:\mathbb{R}^{2n}\times\mathbb{R}^{p}\times\mathbb{R}^{l}\times\mathbb{R}^{l}\rightarrow\mathbb{R}$
is given as
\begin{multline}
\hat{H}\left(\zeta,\hat{\theta},\hat{W}_{c},\hat{W}_{a}\right)=r\left(\zeta,\hat{u}\left(\zeta,\hat{W}_{a}\right)\right)\\
+\frac{\partial\hat{V}\left(\zeta,\hat{W}_{c}\right)}{\partial\zeta}\left(Y_{res}\left(\zeta\right)\hat{\theta}+f_{0_{res}}\left(\zeta\right)+g\hat{u}\left(\zeta,\hat{W}_{a}\right)\right).\label{eq:HJI_approx}
\end{multline}
The error between the optimal and approximate Hamiltonian is called
the Bellman error $\delta:\mathbb{R}^{2n}\times\mathbb{R}^{p}\times\mathbb{R}^{l}\times\mathbb{R}^{l}\rightarrow\mathbb{R}$,
given as
\begin{equation}
\delta\left(\zeta,\hat{\theta},\hat{W}_{c},\hat{W}_{a}\right)=\hat{H}\left(\zeta,\hat{\theta},\hat{W}_{c},\hat{W}_{a}\right)-H\left(\zeta\right),\label{eq:bellman_err}
\end{equation}
where $H\left(\zeta\right)=0\:\forall\zeta\in\mathbb{R}^{2n}$. Therefore,
the Bellman error can be written in a measurable form as
\[
\delta\left(\zeta,\hat{\theta},\hat{W}_{c},\hat{W}_{a}\right)=r\left(\zeta,\hat{u}\left(\zeta,\hat{W}_{a}\right)\right)+\hat{W}_{c}^{T}\omega\left(\zeta,\hat{\theta},\hat{W}_{a}\right),
\]
where $\omega:\mathbb{R}^{2n}\rightarrow\mathbb{R}^{l}$ is given
by 
\[
\omega\left(\zeta,\hat{\theta},\hat{W}_{a}\right)=\sigma'\left(Y_{res}\left(\zeta\right)\hat{\theta}+f_{0_{res}}\left(\zeta\right)+g\hat{u}\left(\zeta,\hat{W}_{a}\right)\right).
\]

The Bellman error may be extrapolated to unexplored regions of the
state space since it depends solely on the approximated system model
and current NN weight estimates. In Section \ref{sec:Stability-Analysis},
Bellman error extrapolation is employed to establish UUB convergence
of the approximate policy to the optimal policy without requiring
persistence of excitation provided the following assumption is satisfied.
\begin{assumption}
\label{thm:concurrent}\cite{Kamalapurkar.Klotz.ea2014a} There exists
a positive constant $\underline{c}$ and set of states $\left\{ \zeta_{k}\in\chi|k=1,2,\ldots,N\right\} $
such that 
\begin{equation}
\inf_{t\in\left[0,\infty\right)}\left[\lambda_{min}\left(\sum_{k=1}^{N}\frac{\omega_{k}\omega_{k}^{T}}{\rho_{k}}\right)\right]=\underline{c},\label{eq:rank_cond}
\end{equation}
where $\omega_{k}\triangleq\omega\left(\zeta_{k},\hat{\theta},\hat{W}_{a}\right)$
and $\rho_{k}\triangleq1+k_{\rho}\omega_{k}{}^{T}\Gamma\omega_{k}$. 
\end{assumption}

The value function least squares update law based on minimization
of the Bellman error is given by 
\begin{multline}
\dot{\hat{W}}_{c}=-\Gamma\left(k_{c1}\frac{\omega\left(\zeta,\hat{\theta},\hat{W}_{a}\right)}{\rho}\delta\left(\zeta,\hat{\theta},\hat{W}_{c},\hat{W}_{a}\right)\right.\\
\left.+\frac{k_{c2}}{N}\sum_{k=1}^{N}\frac{\omega_{k}}{\rho_{k}}\delta_{k}\right),\label{eq:wc_dot}
\end{multline}
\begin{equation}
\dot{\Gamma}=\begin{cases}
\beta\Gamma-k_{c1}\Gamma\frac{\omega\left(\zeta,\hat{\theta},\hat{W}_{a}\right)\omega\left(\zeta,\hat{\theta},\hat{W}_{a}\right)^{T}}{\rho}\Gamma, & \left\Vert \Gamma\right\Vert \leq\overline{\Gamma}\\
0 & \mathrm{otherwise}
\end{cases},\label{eq:gama_dot}
\end{equation}
where $k_{c1},\:k_{c2}\in\mathbb{R}$ are a positive adaptation gains,
$\delta_{k}\triangleq\delta\left(\zeta_{k},\hat{\theta},\hat{W}_{c},\hat{W}_{a}\right)$
is the extrapolated Bellman error, $\left\Vert \Gamma\left(t_{0}\right)\right\Vert =\left\Vert \Gamma_{0}\right\Vert \leq\bar{\Gamma}$
is the initial adaptation gain, $\bar{\Gamma}\in\mathbb{R}$ is a
positive saturation gain, $\beta\in\mathbb{R}$ is a positive forgetting
factor, and 
\[
\rho\triangleq1+k_{\rho}\omega\left(\zeta,\hat{\theta},\hat{W}_{a}\right)^{T}\Gamma\omega\left(\zeta,\hat{\theta},\hat{W}_{a}\right)
\]
 is a normalization constant, where $k_{\rho}\in\mathbb{R}$ is a
positive gain. The update law in (\ref{eq:wc_dot}) and (\ref{eq:gama_dot})
ensures that 
\[
\underline{\Gamma}\leq\left\Vert \Gamma\right\Vert \leq\overline{\Gamma},\:\forall t\in\left[0,\infty\right).
\]
The policy NN update law is given by 
\begin{equation}
\dot{\hat{W}}_{a}=\mbox{proj}\left\{ -k_{a}\left(\hat{W}_{a}-\hat{W}_{c}\right)\right\} ,\label{eq:wa_dot}
\end{equation}
where $k_{a}\in\mathbb{R}$ is an positive gain, and $\mbox{proj}\left\{ \cdot\right\} $
is a smooth projection operator\footnote{See Section 4.4 in \cite{Ioannou1996} or Remark 3.6 in \cite{Dixon2003}
for details of the projection operator.} used to bound the weight estimates. Using properties of the projection
operator, the policy NN weight estimation error can be bounded above
by positive constant.

Using the definition in (\ref{eq:virtual_control}), the force and
moment applied to the vehicle, described in (\ref{eq:dyn}), is given
in terms of the approximated optimal virtual control (\ref{eq:u_approx})
and the compensation term approximation in (\ref{eq:tau_c_approx})
as
\begin{equation}
\hat{\tau}_{b}=\hat{u}\left(\zeta,\hat{W}_{a}\right)+\hat{\tau}_{c}\left(\zeta,\hat{\theta},\nu_{c},\dot{\nu}_{c}\right).\label{eq:tb_hat}
\end{equation}

\section{Stability Analysis\label{sec:Stability-Analysis}}

For notational brevity, all function dependencies from previous sections
will be henceforth suppressed. An unmeasurable form of the Bellman
error can be written using $\eqref{eq:ham}$, $\eqref{eq:HJI_approx}$
and $\eqref{eq:bellman_err}$, as%
\begin{multline}
\delta=-\tilde{W}_{c}^{T}\omega-W^{T}\sigma'Y_{res}\tilde{\theta}-\epsilon'\left(Y_{res}\theta+f_{0_{res}}\right)\\
+\frac{1}{4}\tilde{W}_{a}^{T}G_{\sigma}\tilde{W}_{a}+\frac{1}{2}\epsilon'G\sigma'{}^{T}W+\frac{1}{4}\epsilon'G\epsilon'^{T},\label{eq:unmeas_HJB}
\end{multline}
where $G\triangleq gR^{-1}g^{T}\in\mathbb{R}^{2n\times2n}$ and $G_{\sigma}\triangleq\sigma'G\sigma'{}^{T}\in\mathbb{R}^{l\times l}$
are symmetric, positive semi-definite matrices. Similarly, the Bellman
error at the sampled data points can be written as
\begin{equation}
\delta_{k}=-\tilde{W}_{c}^{T}\omega_{k}-W^{T}\sigma_{k}'\left(Y_{res_{k}}\tilde{\theta}\right)+\frac{1}{4}\tilde{W}_{a}^{T}G_{\sigma k}\tilde{W}_{a}+E_{k},\label{eq:unmeas_HJB_j}
\end{equation}
where 
\[
E_{k}\triangleq\frac{1}{2}\epsilon_{k}'G\sigma_{k}'{}^{T}W+\frac{1}{4}\epsilon_{k}'G\epsilon_{k}'^{T}-\epsilon_{k}'\left(Y_{res_{k}}\theta+f_{0_{res_{k}}}\right)\in\mathbb{R}
\]
 is a constant at each data point, and the notation $F_{k}$ denotes
the function $F\left(\zeta,\cdot\right)$ evaluated at the sampled
state, i.e., $F_{k}\left(\cdot\right)=F\left(\zeta_{k},\cdot\right)$.
The functions $Y_{res}$ and $f_{0_{res}}$ on the compact set $\chi$
are Lipschitz continuous and can be bounded by
\[
\left\Vert Y_{res}\right\Vert \leq L_{Y_{res}}\left\Vert \zeta\right\Vert ,\:\forall\zeta\in\chi,
\]
\[
\left\Vert f_{0_{res}}\right\Vert \leq L_{f_{0res}}\left\Vert \zeta\right\Vert ,\:\forall\zeta\in\chi,
\]
respectively, where $L_{Y_{res}}$ and $L_{f_{0res}}$ are positive
constants.

To facilitate the subsequent stability analysis, consider the candidate
Lyapunov function $V_{L}:\mathbb{R}^{2n}\times\mathbb{R}^{l}\times\mathbb{R}^{l}\times\mathbb{R}^{p}\rightarrow\left[0,\infty\right)$
given as
\[
V_{L}\left(Z\right)=V\left(\zeta\right)+\frac{1}{2}\tilde{W_{c}}^{T}\Gamma^{-1}\tilde{W}_{c}+\frac{1}{2}\tilde{W}_{a}^{T}\tilde{W}_{a}+V_{P}\left(Z_{P}\right),
\]
where $Z\triangleq\left[\begin{array}{cccc}
\zeta^{T} & \tilde{W}_{c}^{T} & \tilde{W}_{a}^{T} & Z_{P}^{T}\end{array}\right]^{T}\in\chi\cup\mathbb{R}^{l}\times\mathbb{R}^{l}\times\mathbb{R}^{p}$. Since the value function $V$ in (\ref{eq:value}) is positive definite,
$V_{L}$ can be bounded by
\begin{equation}
\underline{\upsilon_{L}}\left(\left\Vert Z\right\Vert \right)\leq V_{L}\left(Z\right)\leq\overline{\upsilon_{L}}\left(\left\Vert Z\right\Vert \right)\label{eq:VL_classK}
\end{equation}
using Lemma 4.3 of \cite{Khalil2002} and (\ref{eq:id_classK}), where
$\underline{\upsilon_{L}},\:\overline{\upsilon_{L}}:\left[0,\infty\right)\rightarrow\left[0,\infty\right)$
are class $\mathcal{K}$ functions. Let $\beta\subset\chi\cup\mathbb{R}^{l}\times\mathbb{R}^{l}\times\mathbb{R}^{p}$
be a compact set, and 
\begin{multline*}
\varphi_{\zeta}=\underline{q}-\frac{k_{c1}\sup_{Z\in\beta}\left\Vert \epsilon'\right\Vert \left(L_{Y_{res}}\left\Vert \theta\right\Vert +L_{f_{0_{res}}}\right)}{2}\\
-\frac{L_{Y_{c}}\left\Vert g\right\Vert \left(\left\Vert W\right\Vert \sup_{Z\in\beta}\left\Vert \sigma'\right\Vert +\sup_{Z\in\beta}\left\Vert \epsilon'\right\Vert \right)}{2},
\end{multline*}
\begin{multline*}
\varphi_{c}=\frac{k_{c2}}{N}\underline{c}-\frac{k_{a}}{2}-\frac{k_{c1}\sup_{Z\in\beta}\left\Vert \epsilon'\right\Vert \left(L_{Y_{res}}\left\Vert \theta\right\Vert +L_{f_{0_{res}}}\right)}{2}\\
-\frac{k_{c1}L_{Y}\sup_{Z\in\beta}\left\Vert \zeta\right\Vert \sup_{Z\in\beta}\left\Vert \sigma'\right\Vert \left\Vert W\right\Vert }{2}\\
-\frac{\frac{k_{c2}}{N}\sum_{j=1}^{n}\left(\left\Vert Y_{res_{j}}\sigma_{j}'\right\Vert \right)\left\Vert W\right\Vert }{2},
\end{multline*}
\[
\varphi_{a}=\frac{k_{a}}{2},
\]
\begin{multline*}
\varphi_{\theta}=k_{\theta}\underline{y}-\frac{\frac{k_{c2}}{N}\sum_{k=1}^{N}\left(\left\Vert Y_{res_{k}}\sigma_{k}'\right\Vert \right)\left\Vert W\right\Vert }{2}\\
-\frac{L_{Y_{c}}\left\Vert g\right\Vert \left(\left\Vert W\right\Vert \sup_{Z\in\beta}\left\Vert \sigma'\right\Vert +\sup_{Z\in\beta}\left\Vert \epsilon'\right\Vert \right)}{2}\\
-\frac{k_{c1}L_{Y_{res}}\left\Vert W\right\Vert \sup_{Z\in\beta}\left\Vert \zeta\right\Vert \sup_{Z\in\beta}\left\Vert \sigma'\right\Vert }{2},
\end{multline*}
\begin{multline*}
\kappa_{c}=\sup_{Z\in\beta}\left\Vert \frac{k_{c2}}{4N}\sum_{j=1}^{N}\tilde{W}_{a}^{T}G_{\sigma_{j}}\tilde{W}_{a}+\frac{k_{c1}}{4}\tilde{W}_{a}^{T}G_{\sigma}\tilde{W}_{a}\right.\\
\left.+k_{c1}\epsilon'G\sigma'{}^{T}W+\frac{k_{c1}}{4}\epsilon'G\epsilon'^{T}+\frac{k_{c2}}{N}\sum_{k=1}^{N}E_{k}\right\Vert ,
\end{multline*}
\begin{eqnarray*}
\kappa_{a} & = & \sup_{Z\in\beta}\left\Vert \frac{1}{2}W^{T}G_{\sigma}+\frac{1}{2}\epsilon'G\sigma'^{T}\right\Vert ,
\end{eqnarray*}
\[
\kappa_{\theta}=k_{\theta}d_{\theta},
\]

\[
\kappa=\sup_{Z\in\beta}\left\Vert \frac{1}{4}\epsilon'G\epsilon'^{T}\right\Vert .
\]
When Assumption \ref{thm:concurrent-1} and \ref{thm:concurrent},
and the sufficient gain conditions
\begin{multline}
\underline{q}>\frac{k_{c1}\sup_{Z\in\beta}\left\Vert \epsilon'\right\Vert \left(L_{Y_{res}}\left\Vert \theta\right\Vert +L_{f_{0_{res}}}\right)}{2},\\
+\frac{L_{Y_{c}}\left\Vert g\right\Vert \left(\left\Vert W\right\Vert \sup_{Z\in\beta}\left\Vert \sigma'\right\Vert +\sup_{Z\in\beta}\left\Vert \epsilon'\right\Vert \right)}{2},\label{eq:sc_q}
\end{multline}
\begin{multline}
\underline{c}>\frac{N}{k_{c2}}\left(\frac{k_{c1}\sup_{Z\in\beta}\left\Vert \epsilon'\right\Vert \left(L_{Y_{res}}\left\Vert \theta\right\Vert +L_{f_{0_{res}}}\right)}{2}+\frac{k_{a}}{2}\right.\\
+\frac{k_{c1}L_{Y}\sup_{Z\in\beta}\left\Vert \zeta\right\Vert \sup_{Z\in\beta}\left\Vert \sigma'\right\Vert \left\Vert W\right\Vert }{2}\\
\left.+\frac{\frac{k_{c2}}{N}\sum_{k=1}^{N}\left(\left\Vert Y_{res_{k}}\sigma_{k}'\right\Vert \right)\left\Vert W\right\Vert }{2}\right),\label{eq:sc_c}
\end{multline}
\begin{multline}
\underline{y}>\frac{1}{k_{\theta}}\left(\frac{\frac{k_{c2}}{N}\sum_{k=1}^{N}\left(\left\Vert Y_{res_{k}}\sigma_{k}'\right\Vert \right)\left\Vert W\right\Vert }{2}\right.\\
+\frac{L_{Y_{c}}\left\Vert g\right\Vert \left(\left\Vert W\right\Vert \sup_{Z\in\beta}\left\Vert \sigma'\right\Vert +\sup_{Z\in\beta}\left\Vert \epsilon'\right\Vert \right)}{2}\\
\left.+\frac{k_{c1}L_{Y_{res}}\left\Vert W\right\Vert \sup_{Z\in\beta}\left\Vert \zeta\right\Vert \sup_{Z\in\beta}\left\Vert \sigma'\right\Vert }{2}\right),\label{eq:sc_qy}
\end{multline}
are satisfied, the constant $K\in\mathbb{R}$ defined as 
\[
K\triangleq\sqrt{\frac{\kappa_{c}^{2}}{2\alpha\varphi_{c}}+\frac{\kappa_{a}^{2}}{2\alpha\varphi_{a}}+\frac{\kappa_{\theta}^{2}}{2\alpha\varphi_{\theta}}+\frac{\kappa}{\alpha}}
\]
is positive, where $\alpha\triangleq\frac{1}{2}\min\left\{ \varphi_{\zeta},\varphi_{c},\varphi_{a},\varphi_{\theta},2\underline{k_{\zeta}}\right\} $.
\begin{thm}
Provided Assumptions \ref{thm:neutral}-\ref{thm:concurrent}, the
sufficient conditions (\ref{eq:sc_q})-(\ref{eq:sc_qy}), and 
\begin{equation}
K<\underline{\upsilon_{L}}^{-1}\left(\overline{\upsilon_{L}}\left(r\right)\right)\label{eq:bound}
\end{equation}
are satisfied, where $r\in\mathbb{R}$ is the radius of the compact
set $\beta$, then the policy in $\eqref{eq:u_approx}$ with the NN
update laws in $\eqref{eq:wc_dot}$-$\eqref{eq:wa_dot}$ guarantee
UUB regulation of the state $\zeta$ and UUB convergence of the approximated
policies $\hat{u}$ to the optimal policy $u^{*}$.
\end{thm}
\begin{IEEEproof}
The time derivative of the candidate Lyapunov function is
\begin{multline}
\dot{V}_{L}=\frac{\partial V}{\partial\zeta}\left(Y\theta+f_{0}\right)+\frac{\partial V}{\partial\zeta}g\left(\hat{u}+\hat{\tau}_{c}\right)-\tilde{W}_{c}^{T}\Gamma^{-1}\dot{\hat{W}}_{c}\\
-\frac{1}{2}\tilde{W}_{c}^{T}\Gamma^{-1}\dot{\Gamma}\Gamma^{-1}\tilde{W}_{c}-\tilde{W}_{a}^{T}\dot{\hat{W}}_{a}+\dot{V}_{P}.\label{eq:vl_dot_1}
\end{multline}
Using $\eqref{eq:HJI}$, $\frac{\partial V}{\partial\zeta}\left(Y\theta+f_{0}\right)=-\frac{\partial V}{\partial\zeta}g\left(u^{*}+\tau_{c}\right)-r\left(\zeta,u^{*}\right)$.
Then, 
\begin{multline*}
\dot{V}_{L}=\frac{\partial V}{\partial\zeta}g\left(\hat{u}+\hat{\tau}_{c}\right)-\frac{\partial V}{\partial\zeta}g\left(u^{*}+\tau_{c}\right)-r\left(\zeta,u^{*}\right)\\
-\tilde{W}_{c}^{T}\Gamma^{-1}\dot{\hat{W}}_{c}-\frac{1}{2}\tilde{W}_{c}^{T}\Gamma^{-1}\dot{\Gamma}\Gamma^{-1}\tilde{W}_{c}-\tilde{W}_{a}^{T}\dot{\hat{W}}_{a}+\dot{V}_{P}.
\end{multline*}
Substituting $\eqref{eq:wc_dot}$ and $\eqref{eq:wa_dot}$ for $\dot{\hat{W}}_{c}$
and $\dot{\hat{W}}_{a}$, respectively, yields
\begin{multline*}
\dot{V}_{L}=-\zeta^{T}Q\zeta-u^{*T}Ru^{*}+\frac{\partial V}{\partial\zeta}g\tilde{\tau}_{c}+\frac{\partial V}{\partial\zeta}g\hat{u}-\frac{\partial V}{\partial\zeta}gu^{*}\\
+\tilde{W}_{c}^{T}\left[k_{c1}\frac{\omega}{\rho}\delta+\frac{k_{c2}}{N}\sum_{j=1}^{N}\frac{\omega_{k}}{\rho_{k}}\delta_{k}\right]+\tilde{W}_{a}^{T}k_{a}\left(\hat{W}_{a}-\hat{W}_{c}\right)\\
-\frac{1}{2}\tilde{W}_{c}^{T}\Gamma^{-1}\left[\left(\beta\Gamma-k_{c1}\Gamma\frac{\omega\omega{}^{T}}{\rho}\Gamma\right)\mathbf{1}_{\left\Vert \Gamma\right\Vert \leq\overline{\Gamma}}\right]\Gamma^{-1}\tilde{W}_{c}+\dot{V}_{P}.
\end{multline*}
Using Young's inequality, $\eqref{eq:value_NN}$, $\eqref{eq:u_NN}$,
$\eqref{eq:u_approx}$, $\eqref{eq:unmeas_HJB}$, and $\eqref{eq:unmeas_HJB_j}$
the Lyapunov derivative can be upper bounded as%
\begin{multline*}
\dot{V}_{L}\leq-\varphi_{\zeta}\left\Vert \zeta\right\Vert ^{2}-\varphi_{c}\left\Vert \tilde{W}_{c}\right\Vert ^{2}-\varphi_{a}\left\Vert \tilde{W}_{a}\right\Vert ^{2}-\varphi_{\theta}\left\Vert \tilde{\theta}\right\Vert ^{2}\\
-\underline{k_{\zeta}}\left\Vert \tilde{\zeta}\right\Vert ^{2}+\kappa_{a}\left\Vert \tilde{W}_{a}\right\Vert +\kappa_{c}\left\Vert \tilde{W}_{c}\right\Vert +\kappa_{\theta}\left\Vert \tilde{\theta}\right\Vert +\kappa.
\end{multline*}
Completing the squares, the upper bound on the Lyapunov derivative
may be written as
\begin{multline*}
\dot{V}_{L}\leq-\frac{\varphi_{\zeta}}{2}\left\Vert \zeta\right\Vert ^{2}-\frac{\varphi_{c}}{2}\left\Vert \tilde{W}_{c}\right\Vert ^{2}-\frac{\varphi_{a}}{2}\left\Vert \tilde{W}_{a}\right\Vert ^{2}\\
-\frac{\varphi_{\theta}}{2}\left\Vert \tilde{\theta}\right\Vert ^{2}-\underline{k_{\zeta}}\left\Vert \tilde{\zeta}\right\Vert ^{2}+\frac{\kappa_{c}^{2}}{2\varphi_{c}}+\frac{\kappa_{a}^{2}}{2\varphi_{a}}+\frac{\kappa_{\theta}^{2}}{2\varphi_{\theta}}+\kappa,
\end{multline*}
which can be further upper bounded as
\begin{equation}
\dot{V}_{L}\leq-\alpha\left\Vert Z\right\Vert ,\:\forall\left\Vert Z\right\Vert \geq K>0.\label{eq:uub}
\end{equation}
Using (\ref{eq:VL_classK}), (\ref{eq:bound}), and (\ref{eq:uub}),
Theorem 4.18 in \cite{Khalil2002} is invoked to conclude that $Z$
is uniformly ultimately bounded, in the sense that $\lim\sup_{t\rightarrow\infty}\left\Vert Z\left(t\right)\right\Vert \leq\underline{\upsilon_{L}}^{-1}\left(\overline{\upsilon_{L}}\left(K\right)\right)$.

Based on the definition of $Z$ and the inequalities in (\ref{eq:VL_classK})
and (\ref{eq:uub}), $\zeta,\tilde{W}_{c},\tilde{W}_{a}\in\mathcal{L}_{\infty}$.
Using the fact that $W$ is upper bounded by a bounded constant and
the definition of the NN weight estimation errors, $\hat{W}_{c},\hat{W}_{a}\in\mathcal{L}_{\infty}$.
Using the policy update laws in (\ref{eq:wa_dot}), $\dot{\hat{W}}_{a}\in\mathcal{L}_{\infty}$.
Since $\hat{W}_{c},\hat{W}_{a},\zeta\in\mathcal{L}_{\infty}$ and
$\sigma,\nabla\sigma$ are continuous functions of $\zeta$, it follows
that $\hat{V},\hat{u}\in\mathcal{L}_{\infty}$. From the dynamics
in (\ref{eq:adp_dynamics_varCur}), $\dot{\zeta}\in\mathcal{L}_{\infty}$.
By the definition in (\ref{eq:bellman_err}), $\delta\in\mathcal{L}_{\infty}$.
By the definition of the normalized value function update law in (\ref{eq:wc_dot}),
$\dot{\hat{W}}_{c}\in\mathcal{L}_{\infty}$.
\end{IEEEproof}

\section{Experimental Validation}

Validation of the proposed controller is demonstrated with experiments
conducted at Ginnie Springs in High Springs, FL, USA. Ginnie Springs
is a second-magnitude spring discharging 142 million liters of freshwater
daily with a spring pool measuring 27.4 m in diameter and 3.7 m deep
\cite{Schmidt2004}. Ginnie Springs was selected to validate the proposed
controller because of its relatively high flow rate and clear waters
for vehicle observation. For clarity of exposition\footnote{The number of basis functions and weights required to support a six
DOF model greatly increases from the set required for the three DOF
model. The increased number of parameters and complexity reduces the
clarity of this proof of principal experiment. } and to remain within the vehicle\textquoteright s depth limitations\footnote{The vehicle's Doppler velocity log has a minimum height over bottom
of approximately 3 m that is required to measure water velocity. A
minimum depth of approximately 0.5 m is required to remove the vehicle
from surface effects. With the depth of the spring nominally 3.7 m,
a narrow window of about 20 cm is left operate the vehicle in heave.}, the developed method is implemented on an AUV, where the surge,
sway, and yaw are controlled by the algorithm represented in (\ref{eq:tb_hat}).

\subsection{Experimental Platform}

\begin{figure}
\begin{centering}
\includegraphics[width=1\columnwidth]{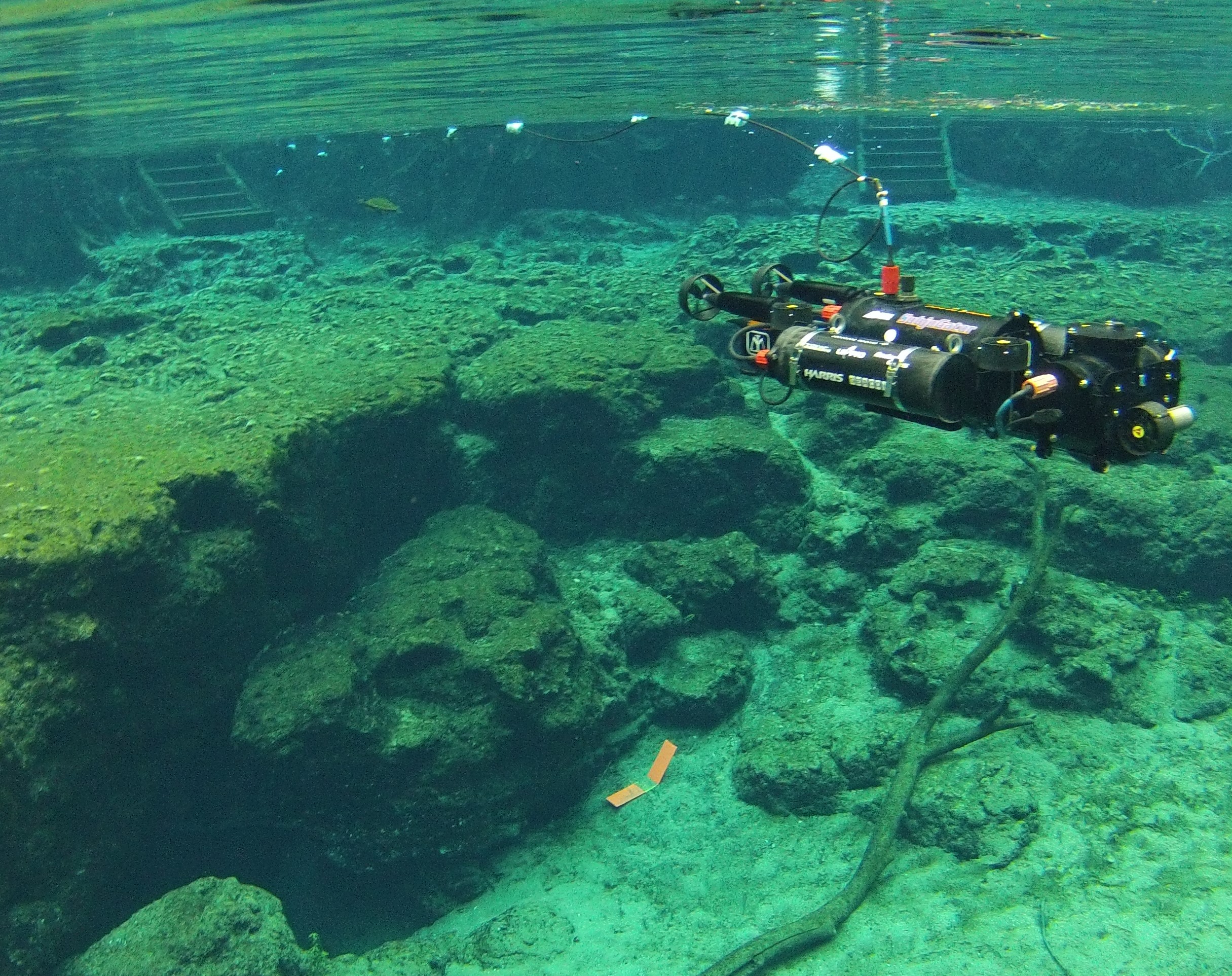}
\par\end{centering}
\caption{\label{fig:SubjuGator-7}SubjuGator 7 AUV operating at Ginnie Springs,
FL.}
\end{figure}

Experiments were conducted on an AUV, SubjuGator 7, developed at the
University of Florida. The AUV, shown in Figure \ref{fig:SubjuGator-7}
, is a small two man portable AUV with a mass of 40.8 kg. The vehicle
is over-actuated with eight bidirectional thrusters.

Designed to be modular, the vehicle has multiple specialized pressure
vessels that house computational capabilities, sensors, batteries,
and mission specific payloads. The central pressure vessel houses
the vehicle's motor controllers, network infrastructure, and core
computing capability. The core computing capability services the vehicles
environmental sensors (e.g. visible light cameras, scanning sonar,
etc.), the vehicles high-level mission planning, and low-level command
and control software. A standard small form factor computer makes
up the computing capability and utilizes a 2.13 GHz server grade quad-core
processor. Located near the front of the vehicle, the navigation vessel
houses the vehicle\textquoteright s basic navigation sensors. The
suite of navigation sensors include an inertial measurement unit,
a Doppler velocity log (DVL), a depth sensor, and a digital compass.
The navigation vessel also includes an embedded 720 MHz processor
for preprocessing and packaging navigation data. Along the sides of
the central pressure vessel, two vessels house 44 Ah of batteries
used for propulsion and electronics.

The vehicle\textquoteright s software runs within the Robot Operating
System framework in the central pressure vessel. For the experiment,
three main software nodes were used: navigation, control, and thruster
mapping nodes. The navigation node receives packaged navigation data
from the navigation pressure vessel where an extended Kalman filter
estimates the vehicle\textquoteright s full state at 50Hz. The controller
node contains the developed controller and system identifier. The
desired force and moment produced by the controller are mapped to
the eight thrusters using a least-squares minimization algorithm in
the thruster mapping node.

\subsection{Controller Implementation}

The implementation of the developed method involves: system identification,
value function iteration, and control iteration. Implementing the
system identifier requires (\ref{eq:ident}), (\ref{eq:para_update}),
and the data set described in Assumption \ref{thm:concurrent-1}.
The data set in Assumption \ref{thm:concurrent-1} was collected in
a swimming pool. The vehicle was commanded to track an exciting trajectory
with a robust integral of the sign of the error (RISE) controller
\cite{Fischer.Hughes.ea2014} while the state-action pairs were recorded.
The recorded data was trimmed to a subset of 40 sampled points that
were selected to maximize the minimum singular value of $\left[\begin{array}{cccc}
Y_{1} & Y_{2} & \ldots & Y_{j}\end{array}\right]$ as in Algorithm 1 of \cite{Chowdhary.Yucelen.ea2012}.

Evaluating the extrapolated Bellman error in (\ref{eq:bellman_err})
with each control iteration is computational expensive. Due to the
limited computational resources available on-board the AUV, the value
function weights were updated at a slower rate (i.e., 5Hz) than the
main control loop (implemented at 50 Hz). The developed controller
was used to control the surge, sway, and yaw states of the AUV, and
a nominal controller was used to regulate the remaining states.

The vehicle uses water profiling data from the DVL to measure the
relative water velocity near the vehicle in addition to bottom tracking
data for the state estimator. By using the state estimator, water
profiling data, and recorded data, the equations used to implement
the proposed controller, i.e., (\ref{eq:ident}), (\ref{eq:para_update}),
(\ref{eq:u_approx}), (\ref{eq:bellman_err}), and (\ref{eq:wc_dot})-(\ref{eq:tb_hat}),
only contain known or measurable quantities.

\subsection{Results}

\begin{figure}
\begin{centering}
\includegraphics[width=1\columnwidth]{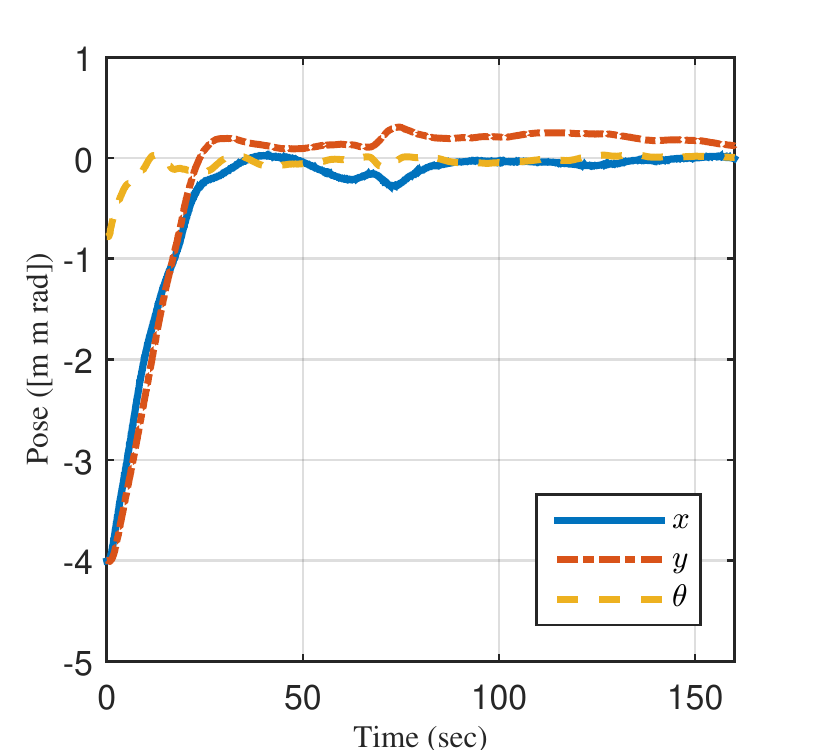}
\par\end{centering}
\begin{centering}
\includegraphics{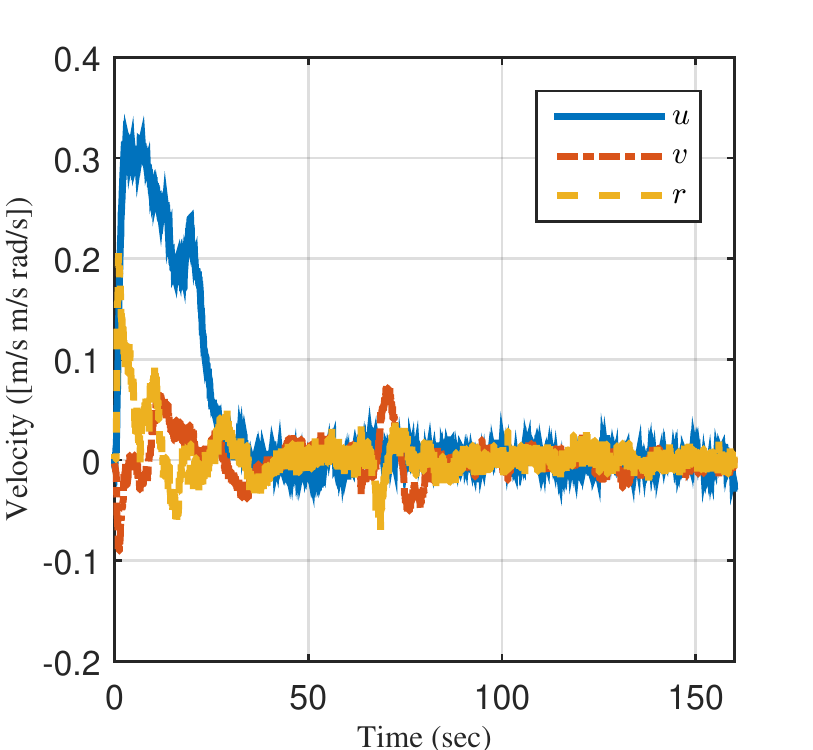}
\par\end{centering}
\caption{\label{fig:error}Inertial position error $\eta$ (top) and body-fixed
velocity error $\nu$ (bottom) of the AUV.}
\end{figure}
\begin{figure}
\begin{centering}
\includegraphics[width=1\columnwidth]{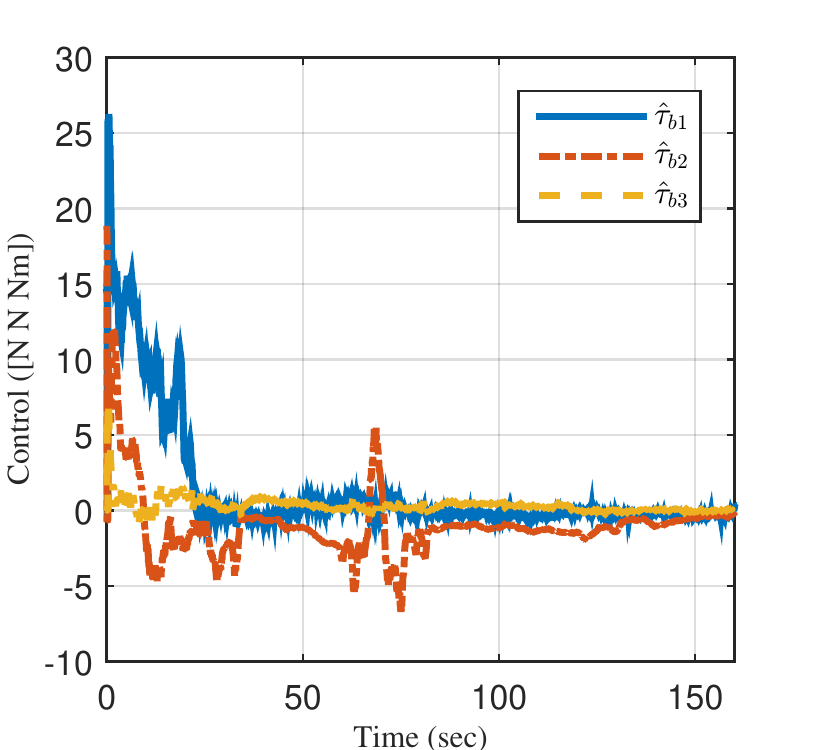}
\par\end{centering}
\caption{\label{fig:control}Body-fixed total control effort $\hat{\tau}_{b}$
commanded about the center of mass of the vehicle.}
\end{figure}
\begin{figure}
\begin{centering}
\includegraphics[width=1\columnwidth]{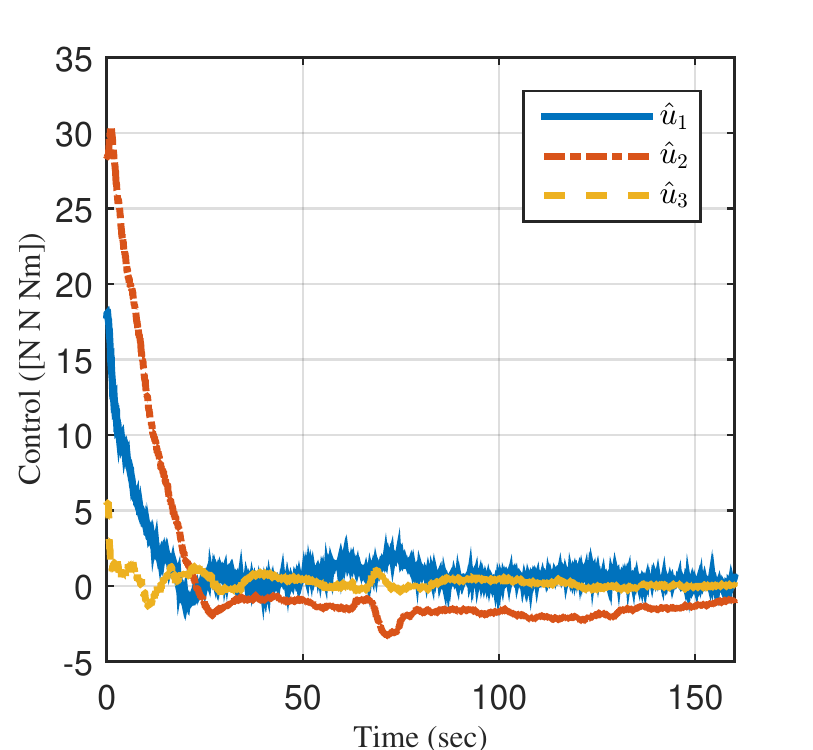}
\par\end{centering}
\caption{\label{fig:optimal-control}Body-fixed optimal control effort $\hat{u}$
commanded about the center of mass of the vehicle.}
\end{figure}
\begin{figure}
\begin{centering}
\includegraphics[width=1\columnwidth]{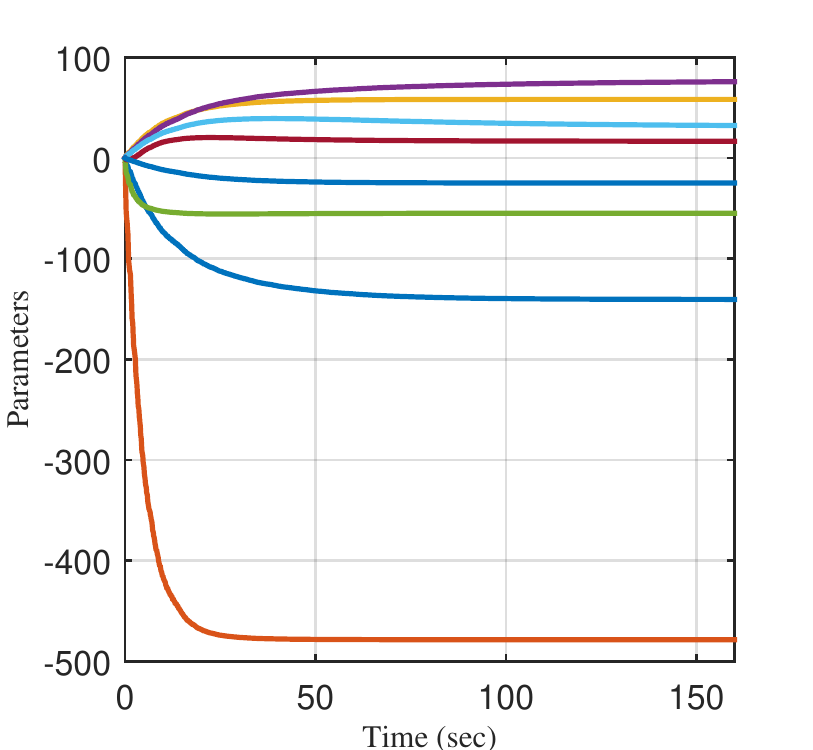}
\par\end{centering}
\caption{\label{fig:params}Identified system parameters determined for the
vehicle online. The parameter definitions may be found in Example
6.2 and Equation 6.100 of \cite{Fossen2011}.}
\end{figure}
\begin{figure}
\begin{centering}
\includegraphics[width=1\columnwidth]{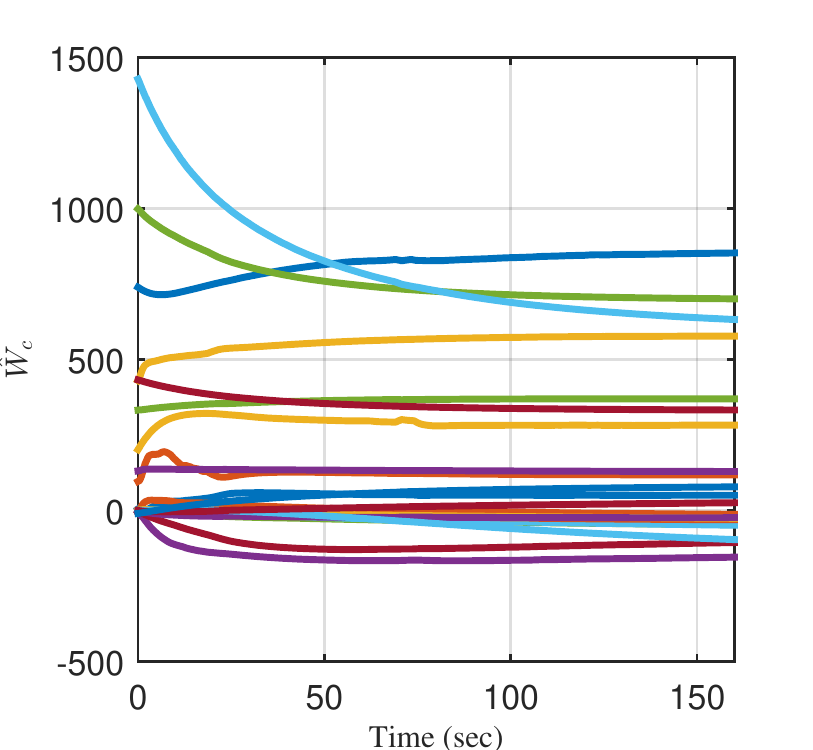}
\par\end{centering}
\begin{centering}
\includegraphics{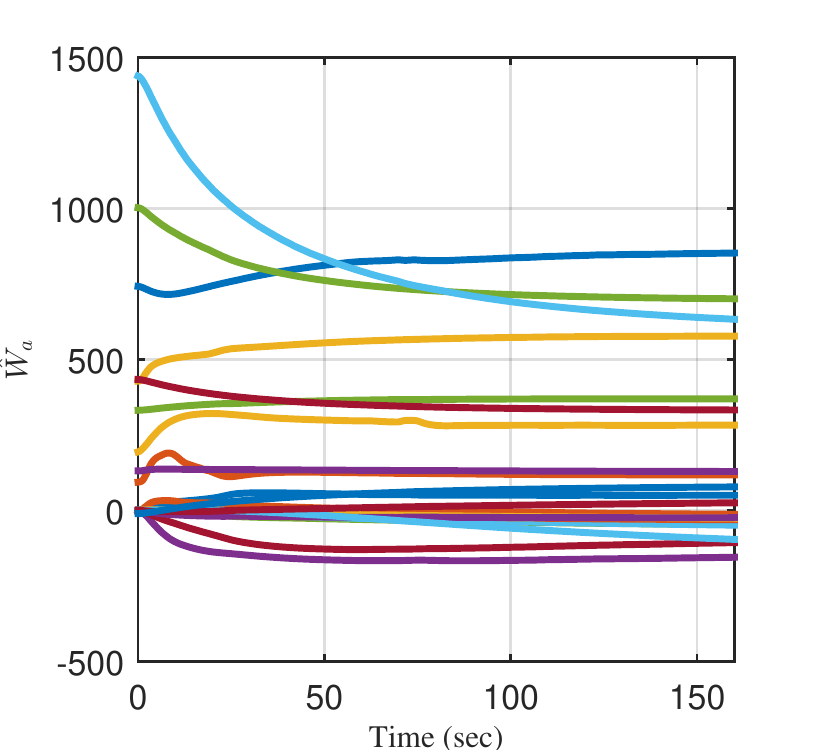}
\par\end{centering}
\caption{\label{fig:weights}Critic $\hat{W}_{c}$ (top) and actor $\hat{W}_{a}$
(bottom) neural network weight estimates online convergence.}
\end{figure}
The vehicle was commanded to hold a station near the vent of Ginnie
Spring. An initial condition of $\zeta\left(t_{0}\right)=\left[\begin{array}{cccccc}
\unit[4]{m} & \unit[4]{m} & \unit[\frac{\pi}{4}]{rad} & \unit[0]{m/s} & \unit[0]{m/s} & \unit[0]{rad/s}\end{array}\right]^{T}$ was given to demonstrate the method's ability to regulate the state.
The optimal control weighting matrices were selected to be $Q=\mathrm{diag}\left(\left[20,50,20,10,10,10\right]\right)$
and $R=I_{3\times3}$. The system identifier adaptation gains were
selected to be $k_{\zeta}=25\times I_{6\times6}$, $k_{\theta}=12.5$,
and $\Gamma_{\theta}=\mbox{diag}\left(\left[187.5,937.5,37.5,37.5,37.5,37.5,37.5,37.5\right]\right)$.
The parameter estimate was initialized with $\hat{\theta}\left(t_{0}\right)=0_{8\times1}$.
The neural network weights were initialized to match the ideal values
for the linearized optimal control problem, which is obtained by solving
the algebraic Riccati equation with the dynamics linearized about
the station. The policy adaptation gains were chosen to be $k_{c1}=0.25$,
$k_{c2}=0.5$, $k_{a}=1$, $k_{p}=0.25$, and $\beta=0.025$. The
adaptation matrix was initialized to $\Gamma_{0}=400\times I_{21\times21}$.
The Bellman error was extrapolated to sampled states that were uniformly
selected throughout the state space in the vehicle\textquoteright s
operating domain.

Figure \ref{fig:error} illustrates the ability of the generated policy
to regulate the state in the presence of the spring's current. Figure
\ref{fig:control} illustrates the total control effort applied to
the body of the vehicle, which includes the estimate of the current
compensation term and approximate optimal control. Figure \ref{fig:optimal-control}
illustrates the output of the approximate optimal policy for the residual
system. Figure \ref{fig:params} illustrates the convergence of the
parameters of the system identifier and Figure \ref{fig:weights}
illustrates convergence of the neural network weights representing
the value function.

The anomaly seen at \textasciitilde{}70 seconds in the total control
effort (Figure \ref{fig:control}) is attributed to a series of incorrect
current velocity measurements. The corruption of the current velocity
measurements is possibly due in part to the extremely low turbidity
in the spring and/or relatively shallow operating depth. Despite presence
of unreliable current velocity measurements the vehicle was able to
regulate the vehicle to its station. The results demonstrate the developed
method's ability to concurrently identify the unknown hydrodynamic
parameters and generate an approximate optimal policy using the identified
model. The vehicle follows the generated policy to achieve its station
keeping objective using industry standard navigation and environmental
sensors (i.e., IMU, DVL).

\section{Conclusion}

The online approximation of an optimal control strategy is developed
to enable station keeping by an AUV. The solution to the HJB equation
is approximated using adaptive dynamic programming. The hydrodynamic
effects are identified online with a concurrent learning based system
identifier. Leveraging the identified model, the developed strategy
simulates exploration of the state space to learn the optimal policy
without the need of a persistently exciting trajectory. A Lyapunov
based stability analysis concludes UUB convergence of the states and
UUB convergence of the approximated policies to the optimal polices.
Experiments in a central Florida second-magnitude spring demonstrates
the ability of the controller to generate and execute an approximate
optimal policy in the presence of a time-varying irrational current.

\section{Acknowledgment}

The authors would like to thank Ginnie Springs Outdoors, LLC, who
provided access to Ginnie Springs for the validation of the developed
controller.

\appendices{}

\section{Extension to constant earth-fixed current\label{sec:Appendix_conCurr}}

In the case where the earth-fixed current is constant, the effects
of the current may be included in the development of the optimal control
problem. The body-relative current velocity $\nu_{c}\left(\zeta\right)$
is state dependent and may be determined from 
\[
\dot{\eta}_{c}=\left[\begin{array}{cc}
\cos\left(\psi\right) & -\sin\left(\psi\right)\\
\sin\left(\psi\right) & \cos\left(\psi\right)
\end{array}\right]\nu_{c},
\]
where $\dot{\eta}_{c}\in\mathbb{R}^{n}$ is the known constant current
velocity in the inertial frame. The functions $Y_{res}\theta$ and
$f_{0_{res}}$ in (\ref{eq:adp_dyn}) can then be redefined as%
\[
Y_{res}\theta\triangleq\left[\begin{array}{c}
0\\
-M^{-1}C_{A}\left(-\nu_{c}\right)\nu_{c}-M^{-1}D\left(-\nu_{c}\right)\nu_{c}\ldots\\
-M^{-1}C_{A}\left(\nu_{r}\right)\nu_{r}-M^{-1}D\left(\nu_{r}\right)\nu_{r}
\end{array}\right],
\]
\[
f_{0_{res}}\triangleq\left[\begin{array}{c}
J_{E}\nu\\
-M^{-1}C_{RB}\left(\nu\right)\nu-M^{-1}G\left(\eta\right)
\end{array}\right],
\]
respectively. The control vector $u$ is 
\[
u=\tau_{b}-\tau_{c}
\]
where $\tau_{c}\left(\zeta\right)\in\mathbb{R}^{n}$ is the control
effort required to keep the vehicle on station given the current and
is redefined as 
\[
\tau_{c}\triangleq-M_{A}\dot{\nu}_{c}-C_{A}\left(-\nu_{c}\right)\nu_{c}-D\left(-\nu_{c}\right)\nu_{c}.
\]

\bibliographystyle{IEEEtran}
\bibliography{master,ncr,encr}

\end{document}